\documentclass[12pt,draftcls,onecolumn,peerreview]{IEEEtran}
\usepackage[cmex10]{amsmath}
\usepackage{amssymb}
\usepackage{cite}
\usepackage{bm}
\usepackage{extarrows}
\usepackage{lineno}
\usepackage{mathrsfs}
\title{Returning to Shannon's Original Meaning}

\author{Xuezhi Yang,~\IEEEmembership{Senior Member,~IEEE }, yangxuezhi@hotmail.com}

\begin{document}
\maketitle


\begin{abstract}
Shannon theory is revisited to show that ergodicity is an indispensable element of channel capacity. The generalized channel capacity $C=\sup_{\bm{X}}\underline{I}(\bm{X}; \bm{Y})$ is checked with a negative conclusion and the popular assertion  "the capacity of a slow fading channel is zero in strict Shannon sense" is found to be conceptually wrong. 

\end{abstract}

\begin{IEEEkeywords}
Channel capacity, Shannon theory, ergodicity. 
\end{IEEEkeywords}

\section{Introduction}

The publication of [1] in 1948 by Shannon marks the birth of information theory. Suppose $X$ and $Y$ are the input and output of a channel,  Shannon's  formula for channel capacity is
\begin{equation}\label{shannonmemless} C=\max_{P_X}I(X;Y),\end{equation} 
where $P_X$ is the distribution of $X$ and the maximum is taken over all possible input distributions $P_X$.

Eq. (1) holds for \emph{memoryless} channels. If a channel has memory, the capacity formula is generalized to  [2][3]
\begin{equation}
\label{capacityblock}
C=\lim_{n\to\infty}\max_{P_{X^n}}\frac{1}{n}I(X^n;Y^n),
\end{equation}
where $X^n =(X_1, X_2, \cdots, X_n)$, $Y^n =(Y_1, Y_2, \cdots, Y_n)$, $P_{X^n}$ is the distribution of $X^n$. It has been well recognized that ergodicity of the channel is an indispensable condition for (1) and (2) to be valid.  

Extending Shannon's definition of channel capacity, Verd$\acute{\textrm{u}}$ and Han  present a general formula for channel capacity, which does not require any assumption such as memorylessness, ergodicity, stationarity, causality, etc [4], as 
\begin{equation}
\label{capacitysup}
C=\sup_{\bm{X}}\underline{I}(\bm{X}; \bm{Y}),
\end{equation}
which will be explained in detail in Section III. 

A slow fading channel where the codeword length is far less than the channel's coherence period is modeled as  non-ergodic. With such a model, instead of \emph{ergodic capacity}, \emph{outage capacity }is used as the metric.  According to the definition in (3), it is widely accepted that the  capacity of a slow fading channel  in  \emph{strict Shannon sense} is zero [5]-[7]. 

However,  the author of  [8] takes exception to the theory in  [4] for the reason that ergodicity is a premise for the definition of channel capacity and therefore it is impossible to define capacity for a non-ergodic channel. In [8],  the capacity of the 1st order Gaussian-Markov process with coherence coefficient $\alpha$  is exploited, and some kind of monotonic behavior is observed with respect to $\alpha$. It is demonstrated in [8]that the capacity of a fading channel changes inversely to the fading rate of the channel, contradicting the  claim that the capacity of a slow fading channel is zero. 

To address the controversies raised by [8],  this paper will discuss such fundamental questions of information theory as: 1. What is Shannon sense? 2. What is developed in [4] and its relationship with Shannon's theory?  3. What is ergodicity and its role in  Shannon's theory? These questions are so fundamental that the answers to them can affect future developments of information theory.

In the sequel of this paper, Shannon sense is discussed in Section II,   Verd$\acute{\textrm{u}}$ and Han's definition is checked in Section III,  ways to cope with slow fading channel in Shannon's framework is discussed in Section IV, and conclusions are given in Section V. 

\section{Exactly, what is Shannon sense?}

Although such saying as "the capacity of a slow fading channel is zero in the strict Shannon sense" is quite popular, there is no explicit definition of Shannon sense, astonishingly.    

Errors happen when messages go through a noisy channel. One can send a message multiple times to decrease the error probability, which however can not be zero no matter how many times the message is sent. Therefore people believed that reliable communication is impossible until Shannon disclosed in [1] that $I(X;Y)$ is a reachable rate with a vanishing error probability. This result  and its converse is  concluded as channel coding theorem, whose proof can be found in any textbook on information theory [2],[9].

First, Shannon assumes the sender knows the channel to the extent of $P_{Y|X}$. Since $I(X;Y)$ is a function of $P_X$ when $P_{Y|X}$ is fixed [9],  then the sender knows the value of $C$ according to (1) and can transmit messages at a rate equal to or lower than $C$ to get an arbitrarily small bit error rate. 

The main techniques used in the proof of channel coding theorem are random codebook and  maximum likelihood or jointly typical decoding algorithm, and the law of large numbers plays a critical role in it.  It is important to notice that, the codeword, channel and output are produced independently in the time domain according to their distributions in the sample space, which bring us to the important concept of \emph{ergodicity}.

Ergodicity is a concept originating from statistical mechanics and thermodynamics. At its most basic level, \emph{ergodicity in mean } refers to the property of a system wherein, over a long period of time,  the time average equals the ensemble average. The ergodicity in Shannon's approach is much stronger than ergodicity in mean, because it is an ergodicity in distribution function. Codewords, channel gains, noises, and etc. are all independently produced in the time domain according to their distribution functions. Then the distributions in time domain are identical to those in ensemble space, which means ergodicity in all orders of moment.

Then we get the key point that  ergodicity is a necessary precondition for channel coding theorem to be valid. Indeed, $I(X;Y)$ is defined in the sample space. If a code in the time domain can achieve that rate, there should be a connection between the time domain and the sample space, which is exactly the concept of ergodicity. Besides, only when the codeword length goes to infinity, can the distributions be precisely reflected in the time domain.  That's why infinite codeword length is required to get a vanishing error probability. 

So, if we detail the question "What is Shannon sense?" as "How does Shannon do to achieve $I(X;Y)$ with a vanishing error probability over an ergodic channel?", the answer would be
\begin{itemize}
  \item The sender knows $P_{Y|X}$;
  \item Random codeword with infinite length; 
  \item Maximum likelihood or jointly typical decoder.
\end{itemize}

To achieve Shannon capacity, both the transmitter and the receiver should have the codebook with  rate $I(X;Y)$ and infinite length. In practices, when the codeword length of a good code is sufficiently large to achieve ergodicity, Shannon capacity  would be a good theory to  predict their behaviors.

\section{Extending Shannon theory to non-ergodic channels?}

As we discussed in the above section, Shanon capacity is a theory about ergodic channels. Can we extend it to non-ergodic channels? That's what  Verd$\acute{\textrm{u}}$ and Han have done in  [4] in which  they give an example  as their motivation in the introduction section:

"Consider a binary channel where the output codeword is equal to the transmitted codeword with probability l/2 and independent of the transmitted codeword with probability l/2. The capacity of this channel is equal to 0 because arbitrarily small error probability is unattainable. However the right-hand side of (2) is equal to l/2 bit/channel use."

It has been proven by Shannon that $I(X;Y)$ is an achievable rate and the conclusion is universally valid in the above-mentioned Shannon sense, so l/2 bit/channel use is quite a normal rate for their example.  Actually, their claim "The capacity of this channel is equal to 0 because arbitrarily small error probability is unattainable" is exactly the naive view about channel capacity before Shannon's work, throwing Shannon's coding away. Based on this observation, there must be something wrong in [4].   Let's  check it out. I just copy the definition of channel capacity in [4], as follows.

"In (3), $\bm{X}$ denotes an input process in the form of a sequence of finite-dimensional distributions.  $\bm{X} = \{X^n =(X_1, X_2, \cdots, X_n)\}_{n=1}^\infty$. We denote by  $\bm{Y} = \{Y^n =(Y_1, Y_2, \cdots, Y_n)\}_{n=1}^\infty$ the corresponding output sequence of the finite-dimensional distributions induced by  $\bm{X}$ via the channel $\bm{W} = \{W^n =P_{Y^n|X^n}: A^n \to B^n\}_{n=1}^\infty$, which is an arbitrary sequence of $n$-dimensional conditional output distributions from $A^n$ to $B^n$, where $A$ and $B$ are the input and output alphabets, respectively. The symbol $\underline{I}(\bm{X}; \bm{Y})$ in (3) is the inf-information rate between $\bm{X}$ and $\bm{Y}$, which is defined in [10] as the \emph{liminf in probability} of the sequence of normalized information densities $(1/n)i_{X^nW^n}( X^n,Y^n)$, where

\begin{equation}
\label{ixnwn}
i_{X^nW^n}(a^n,b^n)=\log\frac{P_{Y^n|X^n}(b^n|a^n)}{P_{Y^n}(b^n)}.
\end{equation}"

The definition of \emph{liminf in probability}   in [10]  is "The  \emph{liminf in probability} of a sequence of random variables $\{{A_n}\}$  is the largest of all  real number $\alpha$, such that for all $\epsilon>0$,  $P[A_n\leqslant\alpha+\epsilon] \to 0$ as $n\to\infty$."

Let's check whether or not (3) is a good extension of Shannon capacity using well known examples. 

\subsection{Memoryless ergodic channel}

Suppose a memoryless ergodic channel with transition probability $P_{Y|X}$, and $X_i, i=1,2,\cdots, n,$ are independent and identically distributed (i.i.d) with a distribution function $P_X$,  then the joint random variables $(X_i,Y_i),i=1,2,\cdots, n,$ are i.i.d with a supposed distribution function $P_{X,Y}$, and  $\log\frac{P_{Y|X}(Y_i|X_i)}{P_{Y}(Y_i)}, i=1,2,\cdots, n$, are i.i.d with mean $I(X;Y)$. Then

\begin{eqnarray}
\frac{1}{n}i_{X^nW^n}(X^n,Y^n) & = & \frac{1}{n}\log\frac{P_{Y^n|X^n}(Y^n|X^n)}{P_{Y^n}(Y^n)}\notag \\
 & = & \frac{1}{n}\sum_{i=1}^n\log\frac{P_{Y|X}(Y_i|X_i)}{P_{Y}(Y_i)}. \notag
\end{eqnarray}

According to the central limit theorem, $\frac{1}{n}i_{X^nW^n}(X^n,Y^n)$  is normally distributed  with mean  $I(X;Y)$ and a diminishing variance as $n\to\infty$. Then 
\begin{eqnarray}
C&=&\sup_{P_{\bm{X}}}\underline{I}(\bm{X}; \bm{Y})\notag\\
&=&\sup_{P_{\bm{X}}}\lim_{n\to \infty} \inf \frac{1}{n}i_{X^nW^n}(X^n,Y^n)\notag\\
&=&\max_{P_X}I(X;Y),\notag
\end{eqnarray}
which is compatible with Shannon's definition. Notice we correct the typo in (3)  as the supremum or  maximum should be  taken over all  possible $P_{\bm{X}}$ instead of $\bm{X}$.

\subsection{Memoryless non-ergodic channel}

Suppose a channel is memoryless and non-ergodic.  At the beginning of each realization of the channel, a random state variable $S$ is randomly chosen to be $s_i, i=1,2.\cdots,K$, which determines the transition  probability to be $P_{Y|X, S}(b|a, s_i), i=1,2.\cdots,K$, which is fixed in the whole process. If  $P_{Y|X, S}(b|a, s_i), i=1,2.\cdots,K$, are different from $P_{Y|X}(b|a)$, then the channel is a non-ergodic one. 

The example as the motivation of [4]  is not clearly described, we guess the authors' intention is to provide a non-ergodic example with two states  "the output codeword is equal to the transmitted codeword" and "the output codeword independent of the transmitted codeword" that fits in the above-mentioned state-based model. Notice example 2 in section VII of [4]  is identical to the motivation example if $\beta=p=1/2$ and fits in this model as well. 

Even though the authors of  [4]  claim (3)  is a general formula, it doesn't have the ability to describe the presented state-based non-ergodic channel, as well as their own examples because they make a conceptual mistake. As we have already leaned, ergodicity is a  relationship of the ensemble space and the time domain, however  (3) doesn't provide the  capability to describe this relationship. With a well-meaning guess, $P_{Y^n|X^n}$ for "the output codeword is equal to the transmitted codeword" is different from  $P_{Y^n|X^n}$ for "the output codeword independent of the transmitted codeword" in the authors' minds throughout their whole paper,  it is an incorrect understanding of the notation $P_{Y^n|X^n}$. The mathematically rigorous understanding of it is 

\begin{equation}
P_{Y^n|X^n}(b^n|a^n)=\sum_{i=1}^K P(S=s_i)P_{Y^n|X^n, S}(b^n|a^n, s_i),
\end{equation}
i.e. $P_{Y^n|X^n}$ is the distribution over all states or the whole ensemble space, not associated to one particular state or realization. 

Based on the above observations, we have the conclusion that, the normalized information density $(1/n)i_{X^nW^n}(X^n,Y^n)$ converges to $I(X;Y)$ for an ergodic channel as well as a non-ergodic one as $n\to\infty$, because their difference  is not reflected in  (4). 

To correct this error and achieve the  objective embodied by the examples in  [4],  $P_{Y^n|X^n}$ in (4) should  be modified to  $P_{Y^n|X^n, S}$, and (4) will be 
\begin{equation}
\label{ }
i_{X^nW^n|S}(a^n,b^n|s_i)=\log\frac{P_{Y^n|X^n, S}(b^n|a^n, s_i)}{P_{Y^n|S}(b^n|s_i)}.
\end{equation}

Then the modified normalized information density will converge to  $I(X;Y|S=s_i)$ when $n\to \infty$, because $P_{Y^n|X^n, S}(b^n|a^n, s_i)$ is exactly  the distribution of the time domain of one specific realization with state $s_i$.  Along with this modification, Eq. (3) should be accordingly modified to 

\begin{equation}
\label{capIXYs}
C=\min_i\sup_{P_{\bm{X}}}\lim_{n\to \infty} \inf \frac{1}{n}i_{X^nW^n|S}(X^n,Y^n|s_i),
\end{equation}
then
\begin{equation}
\label{capIXYsss}
C=\min_i \max_{P_X} I(X;Y|S=s_i).
\end{equation}

According to (8), the capacity of the motivation example is 0, the capacity of example 2 is $1-h(p)$, identical to what is claimed by the authors of [4] and thus achieving their objective.  

\section{Dealing with Non-ergodicity in Shannon's Framework}

Eq. (7) and  (8) is a way to deal with non-ergodicity.  Let us use the first order  Gaussian-Markov  process in [8] which can cover various cases both in theory and practices to  illustrate the basic concepts in information theory.  The channel
\begin{equation}
\label{slowfadingchannel}
Y_i=G_iX_i+Z_i
\end{equation} 
is of first order  Gaussian-Markov  [11] if 

\begin{equation} \label{Gaussian-Markov}G_{i}=\alpha G_{i-1}+\sqrt{1-\alpha^2}W_i,  \end{equation} 
where $i=1,2,\cdots, \infty$, $0\leqslant\alpha\leqslant1$, $G_1$ is Gaussian with variance $\sigma^2_G$,  $\{W_i\}$ is an i.i.d. and Gaussian process with variance $\sigma^2_G$. 

The capacity of such a channel is expressed by (2) in general, which is a Shannon sense one and ergodicity is a necessity.  

In theory, the code length in Shannon's coding is infinite so  all states of the channel are covered to reach ergodicity  for $\alpha<1$, no matter how close $\alpha$ is to 1 and how slow the fading is. That is, in Shanon's sense all channels are fast fading because the code length is  infinite.  Even $\alpha=1$ and $G_i=G_{i-1}$,  the channel can still be theoretically considered as ergodic, deeming all states of the channel can be covered in infinite time.  In practices, the code length should be long enough to cover all states of  channel randomness, including noise and fading to reach full diversity and ergodicity. 

In practices with codeword length much shorter than the coherent period of a fading channel, the channel is modeled as  slow-fading  and non-ergodic.  Typically,  when $\alpha=1$,  the channel gain $G_i$ is Gaussian but fixed for each realization. According to the definition of  (7) and its derivation (8), the capacity is zero, which is the popular view in so many textbooks. But,  (7) is not a Shannon sense capacity, because we have already had the Shannon sense capacity by (2), contradictory to (7). 

After correcting the conceptual error of [4], (8) is technically a valid metric.  But we need to ask, why would we need such a metric? As we have already well known, people used to think reliable communication is not possible until Shannon's theory, which has been guiding the development of communication technology and industry, leading to the discovery of good codes such as Turbo, LDPC and Polar codes. On the contrary, if a theory tells us the capacity of a channel is zero, what does it mean? Does it mean all  messages we receive through that channel are either wrong or fake, because zero semantically means nothing?  If not, such a metric is inappropriate.  The reflection from the practical perspective suggests we should have a deeper think over the concept of ergodicity. 

On a philosophical level, a correct theory should conform to realities. The distribution of a random variable is a theory, it should be identical to the really happening distribution in a realization to be a correct theory, and that's exactly the concept of ergodicity.   When we talk about non-ergodicity, there is actually a gap between theory and practice, which is the essential reason for the bizarre  conclusion of  [4].  For example, if we observe a channel with gain $G_0$, we would use the theory that the channel gain is $G_0$, a constant,  because that's what is really happening, then the channel is an addictive white Gaussian noise (AWGN) channel with gain $G_0$.  It is wrong to regard  $G_0$ as a realization of a random variable and the channel is non-ergodic, because the channel gain doesn't exhibit any randomness.  More generally,  in the state-based model presented in Section III, if $S$ sticks to $s_i$ in a realization, then use $P_{Y^n|X^n, S}(b^n|a^n, s_i) $ instead of $P_{Y^n|X^n}(b^n|a^n)$ as the distribution function and then the channel is ergodic with capacity  $\max_{P_X} I(X;Y|S=s_i)$, assuming the sender knows $P_{Y^n|X^n, S}(b^n|a^n, s_i)$ of course, which is a necessary part of Shannon sense.

In the case of slow fading channel with short codeword, the concept of outage capacity $C_\epsilon$  could be built on basis of Shannon theory.  The time line is divided into many pieces, each of which is modeled in Shannon's framework as an AWGN channel, based on which the channel gain  is further regarded as a realization of a random variable $G$. Notice here, the channel gain is considered to be a constant for each piece, but regarded as a realization of a  random variable when considering all pieces. That is the philosophy of theory and practice.  Suppose  $G$ has  a continuous probability density function, then for $0\leqslant\epsilon\leqslant1$, outage capacity $C_\epsilon$ is the real  number such that

\begin{equation}
\label{outage}
P(\max_{P_X} I(X;Y|G)<C_\epsilon)=\epsilon.
\end{equation}

As we have discussed before,  slow-fading channel is not a Shannon sense concept. All channels are either AGWN or fast fading in Shannon's framework. That is to say, when we cut a slow-fading channel into many AWGN channels, there is a Shannon sense capacity for each of them, but selecting a specific value among them as the metric of the slow fading channel as a whole is not an operation in Shannon's methodology. Once again, Shannon's way to do it is  (2). So $C_\epsilon$ is an engineeringly desired  concept, but not a Shannon sense one. Obviously, under the definition of (11), for a Rayleigh fading channel,  $C_0=0$, $C_1=\infty$, which are of no engineering signifcance.  The typical value for $C_\epsilon$  would be such ones as $C_{0.01}$ and $C_{0.1}$, with $\epsilon$ approximately equivalent to block error rate. 

 \section{conclusions}

Shannon's theory reveals that $I(X;Y)$ is a reachable rate with vanishing bit error rate, and the law of large number based on ergodicity plays a critical role.  Authors of [4] try to extend Shannon's theory to non-ergodic channels, but it is just a wrong understanding of Shannon's theory and the concept of ergodicity.  Shannon's work makes people believe that reliable communication is possible, while  [4]  retro-progresses back to the era before Shannon by implying  "the capacity of a slow fading channel in the strict Shannon sense is zero", which has been misleading researchers and students.  It's time to correct this widely-spread and long-lasting error and return to Shannon's original meaning as a  foundation stone based on which to further develop Shannon's theory.

\bigskip\bigskip\bigskip

\section*{References}

[1] C.~E. Shannon, ``A mathematical theory of communication,'' \emph{Bell System
  Technical Journal}, vol.~27, pp. 379--423 and 623--656, July and Oct. 1948.

[2] R.~G. Gallager, \emph{Information Theory and Reliable Communication}.\hskip 1em
  plus 0.5em minus 0.4em\relax New York: Wiley, 1968.

[3] A.~J. Goldsmith and P.~P. Varaiya, ``Capacity, mutual information, and coding
  for finite-state {Markov} channels,'' \emph{IEEE Trans. Inf. Theory},
  vol.~42, no.~3, pp. 868--886, 1996.

[4] S.~Verd$\acute{\textrm{u}}$ and {Te Sun Han}, ``A general formula for channel
  capacity,'' \emph{IEEE Trans. Inf. Theory}, vol.~40, no.~4, pp. 1147--1157,
  July 1994.

[5] A.~Goldsmith, \emph{Wireless Communications}.\hskip 1em plus 0.5em minus
  0.4em\relax Cambridge University Press, 2005.

[6] D.~Tse and P.~Viswanath, \emph{Fundamentals of Wireless Communication}.\hskip
  1em plus 0.5em minus 0.4em\relax Cambridge Univ. Press, 2005.

[7] E.~Biglieri, J.~Proakis, and S.~Shamai, ``Fading channels:
  information-theoretic and communications aspects,'' \emph{IEEE Trans. Inf.
  Theory}, vol.~44, no.~6, pp. 2619--2692, Oct 1998.

[8] X.~Yang, ``Capacity of fading channels without channel side information,''
  https://arxiv.org/ abs/1903.12360, Mar 2019.

[9 ]T.~M. Cover and J.~A. Thomas, \emph{Elements of Information Theory}.\hskip 1em
  plus 0.5em minus 0.4em\relax Tsinghua University Press, 1991.

[10] T.~S. {Han} and S.~Verd$\acute{\textrm{u}}$, ``Approximation theory of output
  statistics,'' \emph{IEEE Trans. Inf. Theory}, vol.~39, no.~3, pp. 752--772,
  May 1993.

[11] M.~{Medard}, ``The effect upon channel capacity in wireless communications of
  perfect and imperfect knowledge of the channel,'' \emph{IEEE Trans. Inf.
  Theory}, vol.~46, no.~3, pp. 933--946, May 2000.



\section{Correspondence to reviewers' comments, Round 2}

After I finished the correspondence to reviewers' comments, I found a  blind point of knowledge common to  reviewer 1,2 and 4.  What should be known by the sender?

Let's take the AWGN channel $Y=X+Z$ as an example to illustrate.

1. The sender knows every sample of $Z$: Then the sender can code $X=S-Z$, where $S$ is the user signal to be sent. then $Y=S-Z+Z=S$. Then we get a noiseless channel with infinite capacity. But there is no way the sender knows every sample of the noise, therefore this theory is useless. 

2. The sender knows the channel model, signal power $P$ and noise power $N$: Then the channel capacity is $\log(1+P/N)$.  $P$ is its own choice, $N$ can be obtained through some measurement which is engineeringly feasible. 

3. The sender knows nothing about the channel: There is no way to ensure reliable communication, the capacity is zero.  This theory is useless because nobody will use a channel that they know nothing of. 

Therefore, the only surviving theory is $C=\log(1+P/N)$, assuming the sender knows $P$ and $N$. Since so many persons don't know about this, it should be an element of Shannon sense, so I add the following texts in the manuscript

"First, Shannon assumes the sender knows the channel to the extent of $P_{Y|X}$. Since $I(X;Y)$ is a function of $P_X$ when $P_{Y|X}$ is fixed [9],  then the sender knows the value of $C$ according to (1) and can transmit messages at a rate equal to or lower than $C$ to get an arbitrarily small bit error rate. "

Hope this clarification will be helpful to better understand the manuscript.

Reviewer: 1

Comments to the Author
The paper still has many confusions and flaws, and I cannot recommend to accept the paper. Details are provided below.

Q: The author comments in Sec. III that "It has been proven by Shannon that $I(X; Y)$ is an achievable
rate and the conclusion is universally valid in the above mentioned Shannon sense, so l/2 bit/channel use is quite a normal rate for their example." Indeed, this is the case if we evaluate $I(X;Y)$, but the statement that this is capacity requires achievability. How can this rate be achieved in the channel at hand?

A: The rate 1/2  is achieved by the standard Shannon coding. A reminder, maybe you didn't realize that a codeword in Shannon's coding contains infinite number of blocks, each of which may be rate 0 or 1. Authors of [4] don't tell you there is some kind of specialty of the channel at hand, so there should be no doubt on a universally valid conclusion that $I(X; Y)=1/2$ is achievable.

Q: A random realization of the channel can support a rate of 0 or 1 (for codes of infinite length) and the transmitter does not know which case it is. A rate of 1/2 cannot be achieved with vanishing error probability for the case of 0. How can the error probability be made to vanish in this case at a rate of 1/2?

A: You are considering a channel with rate 0, so a rate of 1/2 can not be achieved. It's quite normal. 

If you think $I(X;Y)=1/2$, you are talking about a channel with rate 1/2, and a codeword for this channel contains infinite number of blocks, each of which may be rate 0 or 1.  If a codeword only contains a block of rate 0, then $I(X;Y)=0$.  

Your mistake here is, you think  $I(X;Y)=1/2$, however the codeword for it contains only a rate 0 block. 

Q: This is the main premise of the comment in Q1 (and follow up in Q2) regarding the author's statement in Sec. III.B "At the beginning of each realization of the channel, a random state variable $S$ is randomly chosen to be $s_i; i =1; 2... ;K$, which determines the transition probability to be ... which is fixed in the whole process." In this case, from the point of view of channel coding, the code is designed for a channel $P(Y^n|X^n; s_i)$ for some unknown $s_i$, and not for a channel $P(Y^n|X^n)$.

A: First, authors of [4] do not describe their model clearly, and there are conceptual errors in it. I present a model in Sec. III.B by guessing their objectives from their examples. In my presented model,  channel coding can only be done for $P(Y^n|X^n; s_i)$, because this is the really happening channel.   $P(Y^n|X^n)$ is a mathematical concept that doesn't happen in each realization. 

Q: Note that the authors response in the response letter that "If $P_{Y^n|X^n}(b^n|a^n) = P_{Y^n|X^n;S}(b^n|a^n;si)$, then the process is ergodic. The authors of [4] had a wrong understanding of these concepts?

A: This is a mathematically  rigorous formula for ergodic channel. It's an information provided to readers for better understanding the concept of ergodicity.

Q: in fact contradicts the author's own claim in the first sentence of Sec. III.B "Suppose a channel is memoryless and non-ergodic" which continues with the example quoted above.

A: Please,  I am not saying the supposed channel that "is memoryless and non-ergodic" satisfies $P_{Y^n|X^n}(b^n|a^n) = P_{Y^n|X^n;S}(b^n|a^n; s_i)$, so there is no contradiction. 

Q: From a coding perspective, for such a channel, it is meaningless to code for the channel $P(Y^n|X^n)$ because this channel will basically never occur.

A: Exactly. The principles have been throughly explained in Section IV, both on a philosophical level and with specific examples. Here is the excerpt.

"On a philosophical level, a correct theory should conform to realities. The distribution of a random variable is a theory, it should be identical to the really happening distribution in a realization to be a correct theory, and that's exactly the concept of ergodicity.   When we talk about non-ergodicity, there is actually a gap between theory and practice, which is the essential reason for the bizarre  conclusion of  [4].  For example, if we observe a channel with gain $G_0$, we would use the theory that the channel gain is $G_0$, a constant,  because that's what is really happening, then the channel is an addictive white Gaussian noise (AWGN) channel with gain $G_0$.  It is wrong to regard  $G_0$ as a realization of a random variable and the channel is non-ergodic, because the channel gain doesn't exhibit any randomness.  More generally,  in the state-based model presented in Section III, if $S$ sticks to $s_i$ in a realization, then use $P_{Y^n|X^n, S}(b^n|a^n, s_i) $ instead of $P_{Y^n|X^n}(b^n|a^n)$ as the distribution function and then the channel is ergodic with capacity  $\max_{P_X} I(X;Y|S=s_i)$, assuming the sender knows $P_{Y^n|X^n, S}(b^n|a^n, s_i)$ of course, which is a necessary part of Shannon sense."

Q: Regarding the authors example with the broken phone realization (response to Q3), the answer is yes!

A: I was expecting a No from you, but got a Yes, astonishingly.  You have the right to think so. But does that mean anything to you? When you are downloading at 100Mbps and tell some other guy "My download speed is 0, and I downloaded a movie in 2 minutes", how can you expect that guy to understands you? I believe you should further explain to him, "Actually, the speed of my broken phone is zero, but I am using a good one with 100 Mbps".  Will you feel strange if the guy reply to you "Why did you even mention the broken phone?". How will you reply to that? Would it be "I know you think my download speed is 100 Mbps, but you know what? It's actually zero, because the code that works for all phones has a rate of zero. I am a scientist, I am smarter than you, hahaha!". I could continue, but would stop here. 

Q: If we want to design a code that works for all phones where some phones receive a code which is identical to the transmitted one, and some phones will receive a code which is independent, then the code that works for all phones has a rate of zero.

A: As explained above, it is not appropriate to have such a thought. If you insist on doing so, the proper name for the capacity you defined should have been "the capacity of the world's worst channel", which is useless and not strange to equal zero. 

Q: Also, it wouldn't help to design a code which works for a combination of a working and a broken phone $P(Y^n|X^n)$ because I will only have one of the two phones $P(Y^n|X^n, s_i)$ and and never a combination of the two.

A: Exactly. If you are using only a working phone, then just define the capacity of it. It's silly to  regard your download speed as the speed of the broken phone.  Would you agree?

Q: Regarding the authors response (to Q4) that the author "has the opinion that it an implementation problem when the sender tries to realize transmission at the channel capacity. Whether or not the sender can achieve that doesn't influence the definition of capacity." The reviewer cannot agree with this comment. Achievability is a main ingredient in the definition of capacity. If we define/evaluate an expression but this cannot be achieved by a coding scheme, then we don't call this capacity. Without achievability, the expression that we would call capacity can have no operational meaning, which is necessary from a communication perspective.

A: Even though I have cut out the texts on this, I would like to share my view on this controversy, just for academic interests. Achievability in information theory means, if you transmit at a speed below the channel capacity $C$, then the BER can be arbitrarily small. The calculation method of $C$ is also the key content of information theory. But how the sender know the variables that are needed to calculate $C$, such as channel gain and noise power,  is not discussed in information theory, neither in Shannon's paper nor in any text book on information theory. It is an implementation problem, which can be solved by measurement and feedback. Please refer to the beginning of this correspondence for further explanation. 

Q: Regarding the last comment (Q5), the author correctly comments that in theory we consider infinite codes, but in practice this is relative. However, the paper compares two theories and the author should make the discussion focused on the theory. The assumption that the "infinite" code fits within a coherence period is practical, but does not fit within the discussed theories and hence cannot be used to contradict [4].

A:  "Short codeword length scenario" is a term defined in practices, meaning the length of the codeword is far less than the coherent period of the channel.When we say "short", it means the actually code length is far shorter than the  coherent period.  In theory, the length of  "short code" is also infinite. Wish that won't surprise you as you have realized the relativity of short and long.  A reminder,  you are actually challenging [4], as authors of [4] claim they "present a general formula for channel capacity, which does not require any assumption such as memorylessness, ergodicity, stationarity, causality, etc", meaning there is no case it can not be fitted in.

Reviewer: 2

Comments to the Author
Despite the revision, the author did not address my comments. The author's argument is misleading and contradictory to itself. The reviewer still suggests rejection.

Q1. The author assumes the scenario where the codeword is long enough to experience different channel realizations. There is no problem with such an assumption. The issue here is that the author simply calls such a channel the slow fading channel, which is contradictory to the definition of slow fading. In slow fading, the channel remains unchanged for the whole codeword, which is irrelevant to how the channel is divided or not.

A: I have pointed out in the first round, there is no such thing as "slow fading" in Shannon's approach. I assume you have got that.  Thank you for presenting your understanding of slow fading clearly this round. My following argument is based on your definition of slow fading. 

Q2. The author argues that the channel capacity of the slow fading channel exists. Sure, the instantaneous channel capacity always exists. However, the issue here is that the transmitter does not know the capacity if the channel state information is not available at the transmitter (e.g., the author can consider that the transmitter even does not know the channel distribution). Hence, the transmitter can transmit at a rate larger than the capacity, such that arbitrarily small error probability is no longer achievable. Hence, the capacity is zero in the strict sense. However, the transmitter can still transmit with a non-zero rate given a target error probability, which is (10).

A: First, there is no such thing as "instantaneous channel capacity" in information theory. Capacity is a statistic from infinite number of inputs and outputs, expressed mathematically as $\max_{P_X}I(X;Y)$. A fading channel has instantaneous channel gain $h$ doesn't mean its instantaneous channel capacity is $\log(1+|h|^2\textbf{SNR})$. This is a common misconception that may come from Tse's textbook (section 5.4.1), although I am not sure whether or not he uses the term  "instantaneous channel capacity". Never ever use it again. 

According to your definition of slow fading, a slow fading channel has constant channel gain for the whole codeword (with infinite length), but the transmitter does not know the channel gain and so the capacity.  I kindly remind you the inner contradiction in your comment.  On one hand, you say there is an unknown non-zero capacity, on the other hand, you say the capacity is zero.  The only way you avoid this contradiction is to call the zero capacity another name, e.g.  "capacity of the world's worst channel", which is not a Shannon sense definition. I have throughly explained about this in Section IV, with the excerpt 

"As we have discussed before, slow-fading channel is not a Shannon sense concept. All channels are either AGWN or fast fading in Shannon's framework. That is to say, when we cut a slow-fading channel into many AWGN channels, there is a Shannon sense capacity for each of them, but selecting a specific value among them as the metric of the slow fading channel as a whole is not an operation in Shannon's methodology. Once again, Shannon's way to do it is (2). "

Besides, from the perspective of practices, your theory is useless as it says the capacity of all slow fading channels are zero, if we are still using slow fading channels to transmit messages. Please refer to the following text from Section IV in the manuscript. 

"On the contrary, if a theory tells us the capacity of a channel is zero, what does it mean? Does it mean all messages we receive through that channel are either wrong or fake, because zero semantically means nothing? If not, such a metric is inappropriate. The reflection from the practical perspective suggests we should have a deeper think over the concept of ergodicity."

For the slow fading channel you present, the consistent theory is, it is an AWGN channel with an unknown channel gain and unknown capacity $C$. If the transmitter want to  transmit at $C$ with arbitrarily small bit error rate, do some measurement to learn the channel gain and capacity $C$. If you can not do the measurement for any reason, its capacity is still $C$, just unknown to you, which means you can not leverage it for the transmission, but doesn't mean the capacity is zero.

Q3. The revised part, i.e., Sec. 4, is from [8]. Nothing new is developed.

A: I recommend the reviewer to read this section carefully to correct the misconceptions that come from [4]. 

Reviewer: 3

Comments to the Author
The author has addressed my comment and I support the acceptance of the paper in its current form.

A: The author appreciates your support. 

Reviewer: 4

Comments to the Author
Please see my comments in the attached file.<Review-Paper-Returning to Shannon? s Original Meaning CL2024-0865.pdf>

I. SUMMARY
This paper revisits a general formula for channel capacity established by Verdu and Han in [1] and subsequently questions the well-accepted statement that ?the channel capacity of a slow fading channel in the Shannon sense is equal to zero?. This is a resubmission attempt and the author submitted a response to the reviewers? comments with no substantial changes to the manuscript itself. Although I went through the response to other reviewers, I will not comment on these leaving it to the respective reviewers. However, I can assert that the response to my comment in the first round was not satisfactory to say the least.
I will identify below a few issues that I have with this paper and I hope the author would be able to address
these in the future if the opportunity to revise is granted.

A: I didn't expect you to raise so many questions, since in the last round you only require me to add more steps and explanations for unexperienced readers. 

II. MAJOR COMMENTS
1) The statement "the capacity in Shannon sense" is well defined, understood and accepted in the IT community as the existence of a code with a positive rate whose error probability of error is asymptotically (in the codeword length) arbitrary small. I am not sure this is worth a debate.

A: Your understanding of "the capacity in Shannon sense" is incomplete. The controversy about Shannon sense is whether or not ergodicity is indispensable in the definition of capacity. Authors of  [4] claim they have defined capacity for non-ergodic channels. 

2) It seems that ergodicity is essential to guarantee that the channel capacity can be expressed in terms of mutual information (either for stationary and memoryless channels or for stationary and ergodic channels). That is, ergodicity ensures that the \emph{liminf} in probability of the normalized information density utilized in the general capacity formula established in [1], converges to a non-random quantity.

A: Exactly, even though "stationary and memoryless channels" and "stationary and ergodic channels" are not parallel, which makes me uncomfortable.

3) For the memoryless binary symmetric channel with cross-over probability 1/2 whose capacity is trivial and that the author is contesting in Section III, third paragraph, I think via standard converse steps, one can prove that any relevant code of rate $R$ must satisfy:
\begin{eqnarray}
R & \leqslant & \max_{P_X}I(X;Y)+\epsilon_n \\
   &  \leqslant & 1-H(Z)+ \epsilon_n \\
   & \leqslant & \epsilon_n 
\end{eqnarray}

where $\epsilon_n \to 0$ as $n \to +\infty $ and (2) follows since $Z \sim U(1/2)$. From (3), it is clear that the channel capacity of this channel is trivial. Again, not clear what the author is claiming here.

A: I am claiming that authors of [4] are challenging Shanon's theory. For the example [4] raises, Shannon says the channel capacity is \textsc{clearly} $1/2$, while the authors of [4] claim the capacity is \textsc{clearly} zero, which is a direct confrontation showing [4] is wrong if we admit Shannon is right.  

"The capacity of "memoryless binary symmetric channel with cross-over probability 1/2" is zero" is a common sense without controversy.  For  this channel,  $I(X;Y)=0$, which is implied in your deduction. However, authors of [4] claim the capacity of a channel with $I(X;Y)=1/2$ is zero. This is a direct challenge to Shannon's theory, and of course is wrong. 

So, in the following paragraph, I highlight that $I(X;Y)$ is universally achievable in Shannon's framework. If the counter example [4] raises is valid, then Shannon is wrong. If we admit Shannon is right, then  the authors of [4] don't understand Shannon correctly. 

4) I agree with the derivation in subsection III.A): when ergodicity and the memoryless properties both hold,
the general formula in [1] reduces to the known maximization of mutual information $\max_{p(X)} I (X;Y )$. But how his treatment contradicts [1] is again neither clear nor justified.

A: The question is not clear. I am saying [4] (or [1] in your references) "is compatible with Shannon's definition" in this case, so there is no contradiction. If you mean the typo I pointed out.  I courteously call it a typo, but it's actually a conceptual error. Please check (4) for the definition of information densities, it is a function of $P_X$ and $P_{Y|X}$, but not a function of $X$ and $Y$. These basic knowledges can be found in T. Cover's textbook, Theorem 2.7.4 on page 31 specifically.

5) In subsection III.B), the author criticizes Example 2 in [1] by stating: "The example as the motivation of [4] is not clearly described, we guess the authors' intention is to provide a non-ergodic example with two states "the output codeword is equal to the transmitted codeword" and "the output codeword independent of the transmitted codeword". I am not sure that this is what Example 2 presents. From my understanding, the channel is noiseless with probability $\beta$, or the channel is a standard BSC with cross-over probability $p$, and thus the output is a noisy-version of the input, instead of independent of it as claimed by the author. 

A: You have got it wrong. I am criticizing "the example as the motivation of [4]" presented in the introduction section, not Example 2.  Their relationship is described by the immediate sentence "Notice example 2 in section VII of [4] is identical to the motivation example if $\beta = p = 1/2$, and fits in this model as well." Please notice "the output codeword independent of the transmitted codeword" is a quote from [4].

Q: Subsequently, the broadcast approach described in Example 2 in [1] to maximize the information rate is elegant and has been widely exploited later on in various contexts including the outage setting (see for example [2]). The connection that the author made with channel with states is in my opinion more relevant to the compound channel, although swapping the max and min in (7) is not justified unless it is a typo.

A:  There is no swap here and it is not a typo. But I forgot a step  in the last version, which is added in this version, as the following.  

"Then the modified normalized information density will converge to  $I(X;Y|S=s_i)$ when $n\to \infty$, because $P_{Y^n|X^n, S}(b^n|a^n, s_i)$ is exactly  the distribution of the time domain of one specific realization with state $s_i$.  Along with this modification, Eq. (3) should be accordingly modified to 

\begin{equation}\notag
C=\min_i\sup_{P_{\bm{X}}}\lim_{n\to \infty} \inf \frac{1}{n}i_{X^nW^n|S}(X^n,Y^n|s_i),
\end{equation}
then
\begin{equation}\notag
C=\min_i \max_{P_X} I(X;Y|S=s_i).
\end{equation}"

Please notice, $\min_i$ is later added in alignment with the states introduced, not from $\inf$.  

 $\lim_{n\to \infty} \inf \frac{1}{n}i_{X^nW^n|S}(X^n,Y^n|s_i)=I(X;Y|S=s_i)$ is proved in Section III A. 
 
 $\min_i$ can not be placed after $\max_{P_X}$ as
 
\begin{equation}\notag
C=\max_{P_X} \min_i I(X;Y|S=s_i),
\end{equation}
because the resulting $i$ obtained from $\min_i I(X;Y|S=s_i)$ might be different for different $P_X$, therefore $C$ could be multi-valued and needs further steps of evaluation to converge to a single value.

I don't know what compound channel means here. The conceptual errors in (3) make it unclear what authors of [4] really mean.  After correction of these errors, the resulting (7) and (8)  agree with the examples in [4], which is explained as 

"According to (8), the capacity of the motivation example is 0, the capacity of example 2 is $1-h(p)$, identical to what is claimed by the authors of [4] and thus achieving their objective. "

Q:From an engineering point-of-view, in many situations, we know the set of channel realizations, e.g., $S =\{s_0,s_1\}$, but we don't know which one the channel will take at a given codeword transmission no matter how long the codeword is. Considering that $\max_{p(X)} I(X;Y|S = s_i)$ as the channel capacity as suggested by  the author in Section IV would be naive if the encoder does not know in which state the channel actually is, even if the state is totally known at the receiver (assuming no feedback).

A: If the encoder doesn't know which state the channel is taking, just admit you don't know the channel capacity.  If you want to know,  do some measurements to learn it. Even you define the capacity as the worst one, if you don't know the value of it, you still can not ensure to transmit reliably at a given speed.  The channel capacity doesn't change to zero just because you don't know it. 

For the simplest AWGN channel with capacity $C=\log(1+P/N)$, if you don't know $N$, the capacity is still $C$ which you don't know. It is not zero. Please refer to the arguments at the beginning part of this correspondence for more detailed explanation. 

III. THE DECISION

In the reviewer's opinion, the paper failed to address previous comments and to make a precise and justified point. Therefore, the reviewer recommends a rejection.

REFERENCES

[1] S. Verdu? and T. S. Han, A general formula for channel capacity, IEEE Trans. Inf. Theory, vol. 40, no. 4, pp. 1147?1157, 1994.

[2] E.Biglieri, J.Proakis, and S. Shamai, Fading channels: Information-theoretic and communications aspects, IEEETrans. Inform.Theory,
vol. 44, no. 6, pp. 2619?2692, Oct. 1998.

\section{Correspondence to reviewers' comments, Round 1}

Reviewer: 1

Comments to the Author
The paper intends to reveal some shortcomings of the work of Verdu and Han [4] which defined a general formula for channel capacity. In the attempt to reveal the shortcomings, the author considers a scenario which does not fit well within the assumptions of Verdu and Han, and hence it is not clear how a conclusion can be made in this case. Since this is the main contribution of the paper, I consider this contribution questionable as detailed next. 

Q1: To be more specific, the beginning of Sec. III discusses (3) which is defined for infinite sequences. The scenario considered afterwards is the case of a channel with state, the channel law depends on a state. The author writes "At the beginning of each realization, a random state variable S is randomly chosen to be si ... which determines the transition probability to be $PY|X;si$ ...  which is fixed in the whole process." The question here is, within the framework of [4], how can one define "the beginning of each realization" specifically focusing on "beginning" and "each". If beginning refers to $X_1$ and $Y_1$, then what does "each" mean, bearing in mind that X and Y are infinite sequences which don't end, and hence there is no "each". In such case, the definition of (5) does not fit within the framework of (4) because for infinite sequences $X^n$ and $Y^n$, when s is chosen it sticks forever, and the averaging in (5) is not needed. 

A: $\bm{X} = \{X^n =(X_1, X_2, \cdots, X_n\}_{n=1}^\infty$ is a stochastic  process, having many realizations, such as this one: $\{X_1=x_1, X_2=x_2, \cdots, X_n=x_n\}_{n=1}^\infty$. The "beginning" of such a realization is the moment just before $X_1=x_1$, and I think  "each" is now clear. 

The correspondence  to "the averaging in (5) is not needed" can be found in the immediate following. 

Q2:The author then writes that this "contradicting the subsequent arguments and examples in [4]" without specifying which subsequent arguments and examples are being referred to. In fact, when s is chosen for an infinite sequence, the channel law is $P_{Y^n|X^n;s_i}$, and (4) coincides with (6) because $P_{Y^n|X^n}$ in (4) is in fact $P_{Y^n|X^n;s_i}$$s_i$ sticks forever). So where is the contradiction/confusion? 

A:  This is exactly the misconception in [4]. 

The example as the motivation of [4]  is not clearly described, we guess the authors' intention is to provide a non-ergodic example with two states  "the output codeword is equal to the transmitted codeword" and "the output codeword independent of the transmitted codeword" that fits in the above-mentioned state-based model. Notice example 2 in section VII of [4]  is identical to the motivation example if $\beta=p=1/2$ and fits in this model as well. 

With a well-meaning guess, $P_{Y^n|X^n}$ for "the output codeword is equal to the transmitted codeword" is different from  $P_{Y^n|X^n}$ for "the output codeword independent of the transmitted codeword" in the authors' minds throughout their whole paper,  it is an incorrect understanding of the notation $P_{Y^n|X^n}$. The mathematically rigorous understanding of it is 

\begin{equation}
P_{Y^n|X^n}(b^n|a^n)=\sum_{i=1}^K P(S=s_i)P_{Y^n|X^n, S}(b^n|a^n, s_i).
\end{equation}

If $P_{Y^n|X^n}(b^n|a^n)= P_{Y^n|X^n, S}(b^n|a^n, s_i)$, then the process is ergodic. The authors of [4]  had a wrong understanding of these concepts.
 
Q3: Note that as a result of the framework in [4], the capacity of a channel with state unknown at the transmitter is as in (7) and (8) (see Elgamal and Kim, Network Info Theory), which is a logical consequence of mathematical assumptions and theorems. The statement of the author that this "is meaningless and ridiculous" seems rather subjective and contradicts logical mathematical reasoning. 

A: I have added relevant arguments why it  is "meaningless and ridiculous" in section IV of the new version.  To make the manuscript a better academic paper, I have removed such harsh words, but I would like to raise an example in life to strengthen my point. If you are downloading a movie at a speed of 100Mbps with your phone ( "the output codeword is equal to the transmitted codeword" ) while having a broken phone in your draw ("the output codeword independent of the transmitted codeword"), will you regard your download speed as zero, because of your broken phone?

Q4:The author argues that a proper metric would be $max_X I(X;Y|S=s_i)$, which one may criticize by saying that this can only be achieved if the sender knows $s_i$. The author then defends this metric by writing "If a capacity theory is about a sufficiently large n, there is no reason the transmitter doesn't know the ergodic mode." This actually contradicts the scenario on the left column of the same page by the author, where s is assumed to change in "each" realization, and the channel law as a result is an average with respect to s. One can similarly ask if capacity is for infinite n, then how can s change in "each" realization? In fact, from an operational point of view, the sender might not know s even in an infinite n regime, since there is always a start time $X_1$,$Y_1$, at which the sender does not know what s will be chosen and has to use a codebook that works for all s since the chosen s is unknown (and will stick forever). 

A: This author has reconsidered these arguments on the sender's knowledge about the state of the channel, and has the opinion that it an implementation problem when the sender tries to realize transmission at the channel capacity. Whether or not the sender can achieve that doesn't influence the definition of capacity. So this author cuts out such sayings about sender's knowledge. 

The question "One can similarly ask if capacity is for infinite n, then how can s change in "each" realization?" is still about the concept of realization, which has been explained in the answer to Q1.  

Q5: In the last two paragraphs of Sec III, the author writes "if definition (3) is applied to the short codeword length scenario..." The question here is why would one apply (3) for a short codeword scenario, when it is defined for infinite n? This application does not follow the assumptions of (3). Also, writing "The time line is divided into many pieces..." begs the question: Are they finite length pieces (short codeword)? If yes, then (3) again does not apply, and this cannot be used to contradict [4]. 

A: Whether or not a codeword is short  is a relative concept. In theory, the length of a codeword is infinite. In practice, if the codeword is long enough to approximately fulfill the requirement  of the theory, it can be regarded as infinite.  "The time line is divided into many pieces...", the length of each piece is short compared to the coherent period of the channel. But for each piece, the length is long enough to be regarded as infinite to apply  (3) .

Reviewer: 2

Comments to the Author

Q:In this letter, the author has tried to challenge the definition of outage capacity. However, the argument has too many flaws. In the slow fading channel and assuming no channel state information at the transmitter, a fixed transmission rate R can be larger than the channel capacity C. This leads to an outage even if the blocklength n is sufficiently large since no coding scheme can achieve an arbitrarily small error probability. 

A: This is a conceptual error which has been throughly explained in this version. Though this reviewer doesn't describe the question clearly, I guess the following text in this version is what he means. 

"The time line is divided into many pieces, each of which is modeled in Shannon's framework as an AWGN channel, based on which the channel gain  is further regarded as a realization of a random variable $G$ with a continuous probability density function.  For $0\leqslant\epsilon\leqslant1$, outage capacity $C_\epsilon$ is the real  number such that

\begin{equation}
\label{outage}
P(\max_{P_X} I(X;Y|G)<C_\epsilon)=\epsilon.
\end{equation}"
 
If the block length n is sufficiently large to achieve ergodicity,  an arbitrarily small error probability can be achieved. It is exactly what has been proved by Shannon. 

Q:The author claims that this could be avoided by having the codeword long enough to experience different channel fading states. This is contradictory to the definition of slow fading, where the channel gain remains unchanged for the whole codeword. Hence, the claim is not correct. Besides, the paper does not contain any new insights and contributions. Due to these reasons, the reviewer suggests rejection.

A: In Shannon's definition, there is no slow-fading channel, all fading channels are fast fading,  because the codeword length is infinite. A slow fading channel is defined from the relationship of codeword length and coherent period of the channel, it's an engineering-perspective definition. More detailed argument can be found in Section IV of this version. 

Reviewer: 3

Comments to the Author

Q: I think the main claim of this paper is that the channel ergodicity is indispensable to define the channel capacity. Otherwise, defining the capacity is impossible and hence the well-accepted statement "The capacity of a slow-fading channel is zero in Shannon sense" is problematic since the latter channel is non-ergodic. This claim also contradicts the seminal work by Han and Verdu where a general capacity formula is provided (for quite arbitrary channels). That is, the author is questioning the validity of (3). There is no counter-example that supports the author's claim and hence this paper lacks details supporting its claim and hence why a rejection is justified. 

A: This author points out the conceptual error Han and Verdu made in Section III B, and raises the first order  Gaussian-Markov process as an example to explain why (3) is invalid. Hope this version would make you happy. 

Reviewer: 4

Comments to the Author

This paper revisits the definition of channel capacity as introduced by Shannon, focusing particularly on the role of channel ergodicity. The author examines the implications of non-ergodic channels, a concept which Verdu and Han address with a general formula for channel capacity. However, the author appears to challenge this proposal, especially in scenarios such as slow-fading channels where the codeword length is shorter than the channel's coherence period, resulting in a capacity of zero according to their definition. Instead, metrics like ergodic capacity and outage capacity are suggested for such cases.

In a related work by the same author (Reference 8), it is demonstrated that the claim of zero capacity is not tenable, particularly through the analysis of a first-order Gaussian Markov process channel.

The paper effectively elucidates why Han and Verdu arrived at the zero-capacity assertion, attributing it to the deliberate treatment of the channel as non-ergodic, consequently leading to the conclusion of zero capacity for slow fading channels within the strict Shannon framework.

Upon reviewing this paper and the referenced work (Reference 8), I find the discussion intriguing. The elucidation on the inadequacy of Han's work is not entirely new, but the present paper provides a clearer rationale behind it, enhancing its significance.

However, the following concerns should be addressed:

Q: 1)    While the paper clarifies the issues with Han and Verdu's approach, more detailed explanations are warranted regarding the role of $P_{Y|X}$ in their work. Furthermore, the derivation from Equation (6) to Equation (7) and the subsequent conclusion of zero capacity require further elucidation. Providing additional explanations would be beneficial for readers unfamiliar with the topic.

A: I think  the role of $P_{Y|X}$ is clear since it is a normative concept in the theory of probability. If an input $x_0$ is fed into the channel with $P_{Y|X}(b|a)$, then the distribution of the output is  $P_{Y|X}(b|x_0)$. This relationship is used in the proof of channel coding theorem. Refer to Section 8.7 of Cover's textbook. 

I have added detailed steps and explanations in Section III A, based on which (7) is a direct result of (6).

Q:  2) It would be advantageous to include a simple example demonstrating how Shannon's theory can be applied to determine the outage capacity of a slow-fading channel with short codewords. The brevity of the paper leaves room for such illustrative examples, which would enhance comprehension for readers.
 
A: I have added Section IV using the first order  Gaussian-Markov process as the example to illustrate the relationship between Shannon capacity and outage capacity. 

Q: Overall, while the paper presents a compelling argument against Han and Verdu's zero-capacity claim, addressing the aforementioned concerns would strengthen its accessibility and impact, particularly for non-expert readers.

A: The author appriciates your support.

\end{document}